\documentclass{ifacconf}
\usepackage{cite}
\usepackage{amsmath,amssymb,amsfonts}
\usepackage{algorithmic}
\usepackage{graphicx}
\usepackage{natbib}
\usepackage{textcomp}
\usepackage{xcolor}
\usepackage{mathrsfs}
\usepackage{psfrag}

\newtheorem{definition}{Definition}
\newtheorem{theorem}{Theorem}
\newtheorem{lemma}{Lemma}
\newtheorem{remark}{Remark}

\begin{document}
\begin{frontmatter}

\title{Robust Output Feedback Consensus for Networked Identical Nonlinear Negative-Imaginary Systems\thanksref{footnoteinfo}} 

\thanks[footnoteinfo]{This work was supported by the Australian Research Council under grant DP190102158.\\ Succeeding research results of this paper have been published in \cite{shi2020robustb} and \cite{shi2020robustc}. See Section \ref{sec:succeeding} for more details.}

\author[First]{Kanghong Shi},
\author[First]{Igor G. Vladimirov},
\author[First]{Ian R. Petersen}

\address[First]{School of Engineering, College of Engineering and Computer Science, Australian National University, Canberra, Acton, ACT 2601, Australia \textnormal{(e-mail: kanghong.shi@anu.edu.au, igor.vladimirov@anu.edu.au, ian.petersen@anu.edu.au)}.}

\begin{abstract}
A robust output feedback consensus problem for networked identical nonlinear negative-imaginary (NI) systems is investigated in this paper. Output consensus is achieved by applying identical linear output strictly negative-imaginary (OSNI) controllers to all the nonlinear NI plants in positive feedback through the network topology. First, we extend the definition of nonlinear NI systems from single-input single-output (SISO) systems to multiple-input multiple-output (MIMO) systems and also extend the definition of OSNI systems to nonlinear scenarios. Asymptotic stability is proved for the closed-loop interconnection of a nonlinear NI system and a nonlinear OSNI system under reasonable assumptions. Then, an NI property and an OSNI-like property are proved for networked identical nonlinear NI systems and networked identical linear OSNI systems, respectively. Output feedback consensus is proved for a network of identical nonlinear NI plants by investigating the stability of its closed-loop interconnection with a network of linear OSNI controllers. This closed-loop interconnection is proposed as a protocol to deal with the output consensus problem for networked identical nonlinear NI systems and is robust against uncertainty in the individual system's model.

\end{abstract}


\end{frontmatter}

\section{Introduction}
Negative-imaginary (NI) systems theory was introduced in \cite{lanzon2008stability} and has rapidly attracted interest among control theory researchers (see \cite{petersen2010feedback,xiong2010negative,song2012negative,mabrok2014generalizing,patra2011stability,bhikkaji2011negative,shi2021negative}). NI systems theory can be regarded as a complementary theory to positive-real (PR) systems theory. NI systems theory can deal with systems of relative degrees zero, one and two, while PR systems theory is only applicable to systems of relative degrees zero and one. NI systems theory is typically applied to deal with systems with co-located position sensors and force actuators. NI system theory has achieved success in the control of flexible structures with highly-resonant dynamics (see \cite{cai2010stability}, \cite{mabrok2014generalizing}); e.g., nano-positioning control (see \cite{das2014resonant}, \cite{das2014mimo}, \cite{das2015multivariable}).

A square, real-rational, proper transfer function $G(s)$ is said to be NI if the following conditions are satisfied (\cite{mabrok2014generalizing}): (1) $G(s)$ has no pole in $Re[s]>0$; (2) $\forall \omega>0$ such that $j\omega$ is not a pole of $G(s)$, $j[G(j\omega)-G(j\omega)^*] \geq 0$; (3) If $s=j\omega_0$ with $\omega_0>0$ is a pole of $G(s)$, then it is a simple pole and the residue matrix $K=\lim_{s\to j\omega_0}(s-j\omega_0)jG(s)$ is Hermitian and positive semi-definite; (4) If $s=0$ is a pole of $G(s)$, then $\lim_{s\to 0}s^kG(s)=0$ for all $k\geq 3$ and $\lim_{s\to 0}s^2G(s)$ is Hermitian and positive semi-definite. This definition includes NI systems with free body motion. For a SISO NI system, its phase-lag is in the range $(0,\pi)$. Hence the Nyquist plot of the system's frequency response $G(j\omega)$ is restricted to the lower half of the complex plane for all positive frequencies.

NI systems theory has recently been extended to nonlinear systems (see \cite{ghallab2018extending}). Generally speaking, a system is said to be nonlinear NI if there exists a positive definite storage function such that its time derivative is not greater than the product of the system's input and the time derivative of the system's output. For a nonlinear system with input $u$, state $x$ and output $y$, where $y$ is only a function of $x$, the nonlinear NI property is equivalent to the dissipativity property with a supply rate $w(u,\dot y)=u\dot y$. \cite{ghallab2018extending} extends several existing results in linear NI systems theory to a class of nonlinear systems with the dissipativity property of NI systems. Asymptotic stability is proved for the closed-loop interconnection of a nonlinear NI system and a so-called weak strict nonlinear NI system under several assumptions.

A subclass of NI systems called output strictly negative-imaginary (OSNI) systems was motivated by the notion of output strictly passive (OSP) systems (see \cite{bhowmick2017lti} and \cite{bhowmick2019output}). A square real-rational, proper transfer function $M(s)$ is said to be OSNI if it is stable and there exists a scalar $\delta>0$ such that $j\omega[M(j\omega)-M(j\omega)^*]-\delta \omega^2\check M(j\omega)^*\check M(j\omega)\geq 0$, $\forall \omega \in \mathbb R\cup\{\infty\}$ where $\check M(j\omega)=M(j\omega)-M(\infty)$. The index $\delta$ describes the level of output strictness. The OSNI property of a system with input $u$, output $y$ and a minimal realisation $(A,B,C,D)$ is equivalent to the dissipativity property with a supply rate of $w(u,\dot {\check y})=u^T\dot {\check y}-\delta \left|\dot{\check y}\right|^2$, where $\check y=y-Du$.

A robust output feedback consensus problem for networked NI systems was investigated in \cite{wang2015robust}. Output feedback consensus is considered as the natural convergence of outputs to a common trajectory (not necessarily constant) by the subsystems themselves under the effect of a network connection. Identical strictly negative-imaginary (SNI) controllers are applied to provide positive feedback control to identical NI plants through a network topology. The output consensus problem is solved as an asymptotic stability problem of the networked system. The asymptotic stability of the interconnection of a single NI system and a single SNI system is guaranteed if the DC-gain of these two cascaded systems has its largest eigenvalue being less than unity. In contrast, the asymptotic stability of the networked control system under consideration is achieved when the cascaded DC-gain $P(0)P_s(0)$ of the NI plant and the SNI controller satisfies a condition involving the Laplacian matrix $\mathcal{L}_n$ corresponding to the network: $\lambda_{max}(P(0)P_s(0))<\frac{1}{\lambda_{max} (\mathcal{L}_n)}$, where $\lambda_{max}(\cdot)$ denotes the largest eigenvalue of a matrix.

This paper is motivated by the output feedback consensus problem of networked identical nonlinear NI systems. With the development of nonlinear NI systems theory, the output feedback consensus problem in \cite{wang2015robust} can be investigated directly in nonlinear scenarios. This work differs from the previous work \cite{ghallab2018extending}, \cite{bhowmick2019output} and \cite{wang2015robust} in the following aspects: (a) nonlinear NI systems are now defined for MIMO systems instead of SISO systems only; (b) \cite{bhowmick2019output} only considered linear OSNI systems while this work provides a definition for general nonlinear OSNI systems in terms of their dissipativity properties; (c) instead of a weak strict nonlinear NI system, a nonlinear OSNI system is now considered in closed-loop interconnection with a nonlinear NI system, and asymptotic stability is proved for the closed-loop interconnection; (d) \cite{wang2015robust} provides a protocol for robust output feedback consensus of networked identical linear NI systems while this work provides a protocol to achieve robust output feedback consensus for networked identical nonlinear NI systems.

Notation: The notation in this paper is standard. $\mathbb R$ and $\mathbb C$ denote the fields of real and complex numbers, respectively, and $\mathbb R^{m\times n}$ and $\mathbb C^{m\times n}$ denote the sets of real and complex matrices of dimension $m\times n$ respectively. $A^T$ and $A^*$ denote the transpose and the complex conjugate transpose of matrix $A$. $Re(s)$ and $Im(s)$ denote the real and imaginary parts of $s\in \mathbb C$ respectively. $\mathscr{RH}_\infty^{m\times m}$ denotes the set of real-rational, stable transfer function matrices. $\overline{\cdot}$ denotes a constant value to a given vector or scalar signal. $|\cdot|$ denotes the Euclidean norm of a vector. Given a matrix $A\in \mathbb R^{m\times n}$, $A>0 (<0)$ means $A$ is positive (negative) definite and $A\geq 0 (\leq 0)$ means $A$ is positive (negative) semi-definite. $\mathbb N(A)$ is the null space of $A$. $I_n$ is the $n\times n$ identity matrix and $\mathbf{1}_n$ is the $n\times 1$ vector with all elements being $1$. $diag\{a_1,a_2,\cdots,a_n\}$ is a diagonal matrix with $a_1,a_2,\cdots,a_n$ on its diagonal. $A\otimes B$ denotes the Kronecker product of matrices $A$ and $B$.

Graph Theory Preliminaries: $\mathcal G=(\mathcal V,\mathcal E)$ where $\mathcal V=\{v_1,v_2,\cdots,v_n\}$ and $\mathcal E=\{e_1,e_2,\cdots,e_l\} \subseteq \mathcal V\times \mathcal V$ describes a graph with $n$ nodes and $l$ edges. The adjacency matrix denoted by $\mathcal A = [a_{ij}]\in \mathbb R^{n\times n}$ is defined so that $a_{ii}=0$, and $\forall i\neq j$, $a_{ij}=1$ if $(v_i,v_j)\in \mathcal E$ and $a_{ij}=0$ otherwise. $\mathcal D=diag\{d_1,d_2,\cdots,d_n\}\in\mathbb R^{n\times n}$ denotes the degree matrix where $d_i=\sum_{j=1}^n a_{ij}$ denotes the degree of node $i$. The Laplacian matrix of a graph $\mathcal G$ is given by $\mathcal {L}_n=\mathcal D- \mathcal A$. A sequence of unrepeated edges in $\mathcal E$ that joins a sequence of nodes in $\mathcal V$ defines a path. An undirected graph is connected if there is a path between every pair of nodes.
\section{Preliminaries}
Consider the following general nonlinear system
\begin{align}
    \dot x=&\ f(x,u),\label{eq:nonlinear system state equation}\\
    y=&\ h(x),\label{eq:nonlinear system output equation}
\end{align}
where $f:\mathbb R^p\times\mathbb R\to\mathbb R^p$ is a Lipschitz continuous function and $h:\mathbb R^p\to\mathbb R$ is a class $C^1$ function.
\begin{definition}(\cite{ghallab2018extending})
The system described by (\ref{eq:nonlinear system state equation}) and (\ref{eq:nonlinear system output equation}) is said to be a nonlinear negative-imaginary system if there exists a positive definite storage function $V:\mathbb R^p\to \mathbb R$ of class $C^1$ such that
\begin{equation*}
    \dot V(x(t))\leq u(t)\dot y(t)
\end{equation*}
for all $t \geq 0$.
\end{definition}

\begin{definition}(See also \cite{bhowmick2019output})\label{def:linear OSNI}
Let $M(s)\in \mathscr{RH}_\infty^{m\times m}$. Then, $M(s)$ is said to be output strictly negative-imaginary (OSNI) if there exists a scalar $\delta > 0$ such that
\begin{equation*}
    j\omega[M(j\omega)-M(j\omega)^*]-2\delta \omega^2\check M(j\omega)^*\check M(j\omega)\geq 0
\end{equation*}
$\forall\omega\in\mathbb R\cup\{\infty\}$, where $\check M(j\omega)=M(j\omega)-M(\infty)$. The parameter $\delta>0$ is an index which quantifies the level
of output strictness of a given OSNI system.
\end{definition}

\begin{theorem}(\cite{bhowmick2019output})\label{theorem:linear OSNI dissipativity}
Let $M(s)$ be a causal, square, LTI system described by the state-space equations $\dot x = Ax+Bu$, $x(0) = 0$ and $y = Cx + Du$, where $A$ is Hurwitz, $D = D^T$ and $(A,B,C,D)$ is a minimal realisation. Let the associated transfer function matrix be $M(s)$ and define $\check y = y-Du$. Let $\delta > 0$ be a given scalar. Then, the following statements are equivalent:\\
i) $M(s)$ is OSNI with a level of output strictness $\delta$;\\
ii) There exists a real matrix $Y = Y^T > 0$ such that
$$AY +YA^T+2\delta(CAY)^T(CAY) \leq 0\quad \textnormal{and}\ B = -AYC^T;$$
iii) The realisation $(A,B,C,D)$ is dissipative with respect to the supply rate
$w(u, \dot {\check y}) = u^T\dot {\check y} - \delta |\dot {\check y}|^2$.
\end{theorem}

\section{An Initial NI Stability Result}
In this section, we show that under suitable assumptions, the closed-loop interconnection of a nonlinear NI system and a nonlinear OSNI system is asymptotically stable. First, we give a definition of nonlinear negative-imaginary systems for MIMO systems and a definition of nonlinear output strictly negative-imaginary systems.
\subsection{Definitions of nonlinear MIMO NI systems and nonlinear OSNI systems}
Consider the following general nonlinear system
\begin{align}
    \dot x(t)=&\ f(x(t),u(t)),\label{eq:state equation of nonlinear OSNI}\\
    y(t)=&\ h(x(t)),\label{eq:output equation of nonlinear OSNI}
\end{align}
where $f:\mathbb R^p\times \mathbb R^m \to \mathbb R^p$ is a Lipschitz continuous function and $h:\mathbb R^p \to \mathbb R^m$ is a class $C^1$ function. Now we extend the nonlinear NI systems definition in \cite{ghallab2018extending} from SISO systems to MIMO systems.
\begin{definition}
The system (\ref{eq:state equation of nonlinear OSNI}), (\ref{eq:output equation of nonlinear OSNI}) is said to be a nonlinear negative-imaginary system if there exists a positive definite storage function $V:\mathbb R^p\to \mathbb R$ of class $C^1$ such that
\begin{equation}\label{eq:NI MIMO definition inequality}
    \dot V(x(t))\leq u(t)^T\dot y(t)
\end{equation}
for all $t\geq 0$.
\end{definition}
Let us give a definition for nonlinear OSNI systems based on the dissipativity property.
\begin{definition}\label{def:nonlinear OSNI}
The system (\ref{eq:state equation of nonlinear OSNI}), (\ref{eq:output equation of nonlinear OSNI}) is said to be a nonlinear output strictly negative-imaginary system if there exists a positive definite storage function $V:\mathbb R^p\to\mathbb R$ of class $C^1$ and a constant $\delta>0$ such that
\begin{equation}\label{eq:dissipativity of OSNI}
    \dot V(x(t))\leq u(t)^T\dot y(t) -\delta \left|\dot y(t)\right|^2
\end{equation}
for all $t\geq 0$. In this case, we also say that system (\ref{eq:state equation of nonlinear OSNI}), (\ref{eq:output equation of nonlinear OSNI}) is nonlinear OSNI with degree of output strictness $\delta$.
\end{definition}

\subsection{Asymptotic stability of the closed-loop interconnection of a nonlinear NI system and a nonlinear OSNI system}
\begin{figure}[h!]
\centering
\psfrag{in_0}{\vspace{-0.1cm}$r=0$}
\psfrag{in_1}{\hspace{0.25cm}$u_1$}
\psfrag{y_1}{\hspace{0.1cm}$y_1$}
\psfrag{in_2}{\hspace{0.05cm}$u_2$}
\psfrag{y_2}{\hspace{0.25cm}$y_2$}
\psfrag{H_1}{{$H_1$}}
\psfrag{H_2}{{$H_2$}}
\includegraphics[width=7cm]{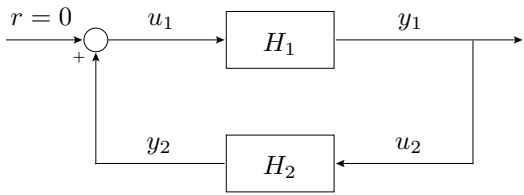}
\caption{Positive feedback interconnection of a nonlinear NI system and a nonlinear OSNI system.}
\label{fig:closed_single}
\end{figure}

In this section, we use the dissipativity property to prove the asymptotic stability of the closed-loop interconnection of a nonlinear NI system and a nonlinear OSNI system, as shown in Fig.~\ref{fig:closed_single}.
\par Consider a nonlinear NI system $H_1$ and a nonlinear OSNI system $H_2$ described as
\begin{align}
  H_i\ (i=1,2):\quad  \dot x_i(t)=&\ f_i(x_i(t),u_i(t)),\label{eq:state-space H_1 state}\\
    y_i(t)=&\ h_i(x_i(t)),\label{eq:state-space H_1 output}
\end{align}
where $f_i:\mathbb R^p\times \mathbb R^m \to \mathbb R^p$ is a Lipschitz continuous function and $h_i:\mathbb R^p \to \mathbb R^m$ is a class $C^1$ function. For the systems $H_1$ and $H_2$, we suppose the following assumptions are satisfied.

\begin{figure}[h!]
\centering
\psfrag{in_1}{$u_1$}
\psfrag{y_1}{$y_1$}
\psfrag{in_2}{$\ u_2$}
\psfrag{y_2}{$y_2$}
\psfrag{H_1}{$H_1$}
\psfrag{H_2}{$H_2$}
\includegraphics[width=8cm]{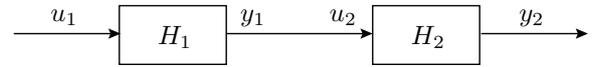}
\caption{Open-loop interconnection of a nonlinear NI system and a nonlinear OSNI system.}
\label{fig:open-loop interconnection of single NI and OSNI}
\end{figure}

For all $x_i(t),u_i(t),y_i(t)$ that satisfy the state equations (\ref{eq:state-space H_1 state}), (\ref{eq:state-space H_1 output}), we assume:\\
Assumption I: Over any time interval $[t_a,t_b]$ where $t_b>t_a$, $y_i(t)$ remains constant if and only if $x_i(t)$ remains constant; i.e., $\dot x_i(t) \equiv 0 \iff \dot y_i(t) \equiv 0$. Moreover, $x_i(t) \equiv 0 \iff y_i(t) \equiv 0$.\\
Assumption II: Over any time interval $[t_a,t_b]$ where $t_b>t_a$, $x_i(t)$ remains constant only if $u_i(t)$ remains constant; i.e., $x_i(t)\equiv \bar x_i \implies u_i(t)\equiv\bar u_i$. Moreover, $x_i(t)\equiv 0\implies u_i(t)\equiv 0$.\\
Assumption III: Given any constant input $\bar u_1$ for the system $H_1$, we obtain a corresponding output $y_1(t)$. Set the input of the system $H_2$ as $u_2(t)\equiv y_1(t)$ in the open-loop interconnection shown in Fig.~\ref{fig:open-loop interconnection of single NI and OSNI}. If the corresponding output of the system $H_2$ is constant; i.e., $y_2(t)\equiv \bar y_2$, then there exists a constant $\gamma\in(0,1)$ such that for any $\bar u_1$ and its corresponding $\bar y_2$, we have
\begin{equation}\label{eq:DC gain condition in SISO AS theorem}
    \bar u_1^T\bar y_2 \leq \gamma |\bar u_1|^2.
\end{equation}
\vspace{-0.5cm}
\begin{theorem}\label{theorem:AS of single interconnection}
Consider the closed-loop positive feedback interconnection of the systems $H_1$ and $H_2$ as shown in Fig.~\ref{fig:closed_single} such that Assumptions I-III are satisfied. If the storage function defined as
\begin{equation*}
    W(x_1,x_2)=V_1(x_1)+V_2(x_2)-h_1(x_1)^Th_2(x_2)
\end{equation*}
is positive definite, where $V_1(x_1)$ and $V_2(x_2)$ are positive definite storage functions that satisfy the nonlinear NI inequality (\ref{eq:NI MIMO definition inequality}) for $H_1$ and the nonlinear OSNI inequality (\ref{eq:dissipativity of OSNI}) for $H_2$, respectively, and $h_1(x_1)$ and $h_2(x_2)$ are the output functions of systems $H_1$ and $H_2$, respectively, then the closed-loop interconnection of the systems $H_1$ and $H_2$ is asymptotically stable.
\end{theorem}
\begin{pf}
Applying Lyapunov's direct method, we take the time derivative of the function $W(x_1,x_2)$. According to (\ref{eq:NI MIMO definition inequality}) and (\ref{eq:dissipativity of OSNI}), we obtain:
\begin{equation}\label{eq:dot W for single closed-loop interconnection}
    \dot W(x_1,x_2)=\dot V_1(x_1)+ \dot V_2(x_2) - u_1^T\dot y_1-u_2^T\dot y_2\leq -\delta \left|\dot y_2\right|^2 \leq 0.
\end{equation}
This inequality shows that the system is Lyapunov stable. We will prove in the following that the equilibrium point at $(x_1,x_2)=(0,0)$ is asymptotically stable.

According to (\ref{eq:dot W for single closed-loop interconnection}), $\dot W(x_1,x_2)=0$ can only hold when $\dot y_2=0$. In other words, $\dot W(x_1,x_2)$ cannot remain zero unless $\dot y_2$ remains zero. According to Assumptions I and II, $\dot y_2(t)\equiv 0\implies\dot x_2(t)\equiv 0\implies \dot u_2(t)\equiv 0$. Due to the feedback interconnection of the systems $H_1$ and $H_2$, it follows that $u_2(t)\equiv y_1(t)$. Hence, we obtain $\dot y_1(t)\equiv 0\implies\dot x_1(t)\equiv 0\implies\dot u_1(t)\equiv 0$. Therefore, the closed-loop system is in steady-state. We let the constant inputs, outputs and states of the systems $H_1$ and $H_2$ be denoted as $\bar u_1$, $\bar u_2$, $\bar y_1$, $\bar y_2$, $\bar x_1$ and $\bar x_2$, respectively. Considering the feedback interconnection, we must have $\bar y_2=\bar u_1$ and therefore inequality (\ref{eq:DC gain condition in SISO AS theorem}) implies
\begin{equation*}
    \bar u_1^T\bar y_2=\left|\bar u_1\right|^2 \leq \gamma \left|\bar u_1\right|^2.
\end{equation*}
This can only hold when $\bar y_2=\bar u_1=0$. According to Assumptions I and II, $\bar y_2=0\implies \bar x_2= 0\implies \bar u_2=0\implies \bar y_1=0\implies \bar x_1=0$.
Therefore, $\dot W(x_1,x_2)$ cannot remain zero unless $(x_1,x_2)=(0,0)$. Hence, $W(x_1,x_2)$ will keep decreasing until $x_1=x_2=0$. According to LaSalle's invariance principle, it follows that the equilibrium point of the closed-loop interconnection of $H_1$ and $H_2$ at $(x_1,x_2)=(0,0)$ is asymptotically stable. This completes the proof.
\end{pf}
\section{Stability of Networked NI Systems}
In this section, we extend Theorem \ref{theorem:AS of single interconnection} from the closed-loop interconnection of a nonlinear NI system and a nonlinear OSNI system to the closed-loop interconnection of networked nonlinear NI systems connected by linear OSNI systems. This system setting is also proposed as a protocol for the robust output feedback consensus of networked identical nonlinear NI systems. To extend Theorem \ref{theorem:AS of single interconnection} to networked systems, we first prove the nonlinear NI property for networked identical nonlinear NI systems, and an OSNI-like property for networked identical linear OSNI controllers. First, we investigate the OSNI property of networked identical linear OSNI controllers by decomposing the entire network into edge-linked pairs of nodes. The following subsection establishes the OSNI property of two connected linear OSNI systems.
\subsection{OSNI property of two connected OSNI systems}
\begin{figure}[h!]
\centering
\psfrag{1}{\hspace{-0.02cm}\large$1$}
\psfrag{2}{\hspace{0cm}\large$2$}
\includegraphics[width=3cm]{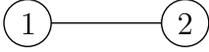}
\caption{An undirected and connected graph consisting of two nodes.}
\label{fig:2_nodes}
\end{figure}
\begin{lemma} 
Let $M(s)$ be a square transfer function matrix described by the state-space representation $\dot x =Ax+Bu$, $y=Cx$, where $u\in\mathbb{R}^m$, $x\in\mathbb{R}^q$, $y\in\mathbb{R}^m$; $A\in\mathbb{R}^{q\times q}$, $B\in\mathbb{R}^{q\times m}$, $C\in\mathbb{R}^{m\times q}$, and $(A,B,C)$ is a minimal realisation of $M(s)$. Consider two identical systems with transfer function $M(s)$ simply connected by the graph described by the Laplacian matrix $\mathcal{L}_2=\left[\begin{matrix}1&-1\\-1&1\end{matrix}\right]$, as shown in Fig.~\ref{fig:2_nodes}. Suppose $M(s)$ is an OSNI transfer function matrix with a level of output strictness $\delta >0$. Then the networked system with transfer function matrix $\mathcal{L}_2\otimes M(s)$ is also OSNI with a level of output strictness $\frac{\delta}{2}$.
\end{lemma}
\begin{pf}
According to Theorem 1, the realisation $(A,B,C)$ is dissipative with respect to the supply rate $\omega(u,\dot y) = u^T \dot y-\delta |\dot y|^2$. Hence, there exists a positive storage function $V_2(x)$ that satisfies $\dot V_2(x) \leq u^T \dot y-\delta |\dot y|^2$. We show in the following that there is also a similar dissipativity property for system corresponding to the transfer function matrix $\mathcal{L}_2\otimes M(s)$.

Let the states, inputs and outputs of the two identical systems with transfer function matrix $M(s)$ be denoted as $x_1,x_2$; $u_1,u_2$ and $y_1,y_2$, respectively. Then the following state equations will be satisfied:
\begin{equation*}
    \begin{aligned}
    \textnormal{System} \ i\ (i=1,2):\quad \dot x_i =&\ Ax_i+Bu_i,\\
    y_i =&\ Cx_i,\\
    \end{aligned}
\end{equation*}
where $(A,B,C)$ defines a minimal realisation for $M(s)$. Let us define the following quantities:
$\Delta x_{ij}=x_i-x_j$, $\Sigma x_{ij}=x_i+x_j$, $\Delta u_{ij}=u_i-u_j$, $\Sigma u_{ij}=u_i+u_j$ and $\Delta y_{ij}=y_i-y_j$.\\
For the networked system with transfer function matrix $$\mathcal{L}_2\otimes M(s)=\left[\begin{matrix}M(s)&-M(s)\\-M(s)&M(s)\end{matrix}\right],$$ we define the input $\tilde u=\left[\begin{matrix}u_1 \\ u_2\end{matrix}\right]$ and the output $\tilde y = \left[\begin{matrix}y_1-y_2 \\ y_2-y_1\end{matrix}\right]=\left[\begin{matrix}\Delta y_{12}\\ \Delta y_{21}\end{matrix}\right]=\left[\begin{matrix}\Delta y_{12}\\ -\Delta y_{12}\end{matrix}\right]$. Also, we define the states of the networked system as $\tilde x =\left[\begin{matrix}x_1-x_2 \\ x_1+x_2\end{matrix}\right]= \left[\begin{matrix}\Delta x_{12} \\ \Sigma x_{12}\end{matrix}\right]$.
Hence, a state-space realisation of the transfer function matrix $\mathcal{L}_2\otimes M(s)$ can be written as follows.
\begin{align}
\dot {\tilde x} =&\left[\begin{matrix}\Delta \dot x_{12} \\ \Sigma \dot x_{12}\end{matrix}\right] = \left[\begin{matrix}A&0\\0&A\end{matrix}\right]\left[\begin{matrix}\Delta x_{12} \\ \Sigma x_{12}\end{matrix}\right]+\left[\begin{matrix}B&0\\0&B\end{matrix}\right]\left[\begin{matrix}\Delta u_{12} \\ \Sigma u_{12}\end{matrix}\right],\label{eq:state-space of 2 nodes state}\\
    \tilde y =& \left[\begin{matrix}\Delta y_{12}\\ -\Delta y_{12}\end{matrix}\right] =\left[\begin{matrix}C&0\\-C& 0\end{matrix}\right]\left[\begin{matrix}\Delta x_{12} \\ \Sigma x_{12}\end{matrix}\right].\label{eq:state-space of 2 nodes output}
\end{align}
Here, the input $\tilde u=\left[\begin{matrix}u_1 \\ u_2\end{matrix}\right]=\left[\begin{matrix}\frac{1}{2}&\frac{1}{2} \\-\frac{1}{2} &\frac{1}{2}\end{matrix}\right]\left[\begin{matrix}\Delta u_{12}\\ \Sigma u_{12}\end{matrix}\right]$. By introducing this transfer function matrix, we arrange the system's state equations as two independent state equations. The first row of the state equation (\ref{eq:state-space of 2 nodes state}) can be considered as a new system together with the first row of the output equation (\ref{eq:state-space of 2 nodes output}). This system has the same state equations as Systems $1$ and $2$, which is a minimal realisation of $M(s)$. The state of the second row in the state equations (\ref{eq:state-space of 2 nodes state}), (\ref{eq:state-space of 2 nodes output}) is unobservable. Since we apply $M(s)$ later as a control law to eliminate the difference between the outputs of edge-linked plants, we are only interested in the difference between the two controllers. Therefore, we split off the first row of the state-space model (\ref{eq:state-space of 2 nodes state}), (\ref{eq:state-space of 2 nodes output}) as a new system, which has the state-space model
\begin{align}
\Delta \dot x_{12} = &\ A\Delta x_{12}+B\Delta u_{12},\label{eq:state-space difference state}\\
\Delta y_{12} =&\ C\Delta x_{12}.\label{eq:state-space difference output}
\end{align}
This system describes the difference of the two networked subsystems and is an OSNI system. However, the input and output of this system are modified versions of the inputs and outputs of the original system with transfer function matrix $\mathcal{L}_2\otimes M(s)$. The OSNI property of transfer function matrix $\mathcal{L}_2 \otimes M(s)$ is established in the following by using the equivalence of the OSNI property and the corresponding dissipativity property.

We take $V_2(\Delta x_{12})$ as the storage function for system (\ref{eq:state-space difference state}), (\ref{eq:state-space difference output}) with transfer function matrix $\mathcal L_2 \otimes M(s)$. By the virtue of the transfer function matrix $M(s)$, for the system (\ref{eq:state-space difference state}), (\ref{eq:state-space difference output}), we have
\begin{equation}\label{eq:dot V_2_{12}}
    \dot V_2(\Delta x_{12})\leq \Delta u_{12}^T\Delta\dot y_{12}-\delta|\Delta \dot y_{12}|^2.
\end{equation}
Though the input $\tilde u$ and output $\tilde y$ of system with transfer function matrix $\mathcal{L}_2\otimes M(s)$ do not appear in (\ref{eq:dot V_2_{12}}), we can replace the terms $\Delta u_{12}^T\Delta\dot y_{12}$ and $|\Delta \dot y_{12}|^2$ with the terms $\tilde u^T \dot {\tilde y}$ and $\frac{1}{2}|\dot {\tilde y}|^2$, respectively, according to the following calculation:
\begin{equation}\label{eq:uy dot}
\begin{aligned}
\tilde u^T\dot{\tilde y}=&\left[\begin{matrix}u_1^T & u_2^T\end{matrix}\right]\left[\begin{matrix}\dot y_1-\dot y_2 \\ \dot y_2-\dot y_1\end{matrix}\right]\\
=&\ u_1^T(\dot y_1-\dot y_2)+u_2^T(\dot y_2 -\dot y_1)\\
=&\ (u_1-u_2)^T(\dot y_1-\dot y_2)\\
=&\ \Delta u_{12}^T\Delta\dot y_{12};
\end{aligned}
\end{equation}
\begin{equation}\label{eq:y dot squared}
\begin{aligned}
    |\dot {\tilde y}|^2=&\left[\begin{matrix}\dot y_1^T-\dot y_2^T & \dot y_2^T-\dot y_1^T\end{matrix}\right]\left[\begin{matrix}\dot y_1-\dot y_2 \\ \dot  y_2-\dot y_1\end{matrix}\right]\\
    =&\ 2|\dot y_1-\dot y_2|^2\\
    =&\ 2|\Delta \dot y_{12}|^2.
\end{aligned}
\end{equation}
Substituting (\ref{eq:uy dot}) and (\ref{eq:y dot squared}) into (\ref{eq:dot V_2_{12}}), we have
\begin{equation*}
    \dot V_2(\Delta x_{12})\leq\tilde u^T \dot {\tilde y}-\frac{1}{2}\delta|\dot {\tilde y}|^2.
\end{equation*}
Therefore, the system (\ref{eq:state-space difference state}), (\ref{eq:state-space difference output}) with transfer function matrix $\mathcal{L}_2\otimes M(s)$ is dissipative with a supply rate $\omega(\tilde u, \dot {\tilde y})=\tilde u^T \dot {\tilde y}-\frac{1}{2}\delta |\dot {\tilde y}|^2$, which implies $\mathcal{L}_2\otimes M(s)$ is OSNI with a level of output strictness $\frac{\delta}{2}$.
\end{pf}
\subsection{An OSNI-like property for a network of OSNI systems}
It is shown above that a system consisting of two identical linear OSNI systems connected by an undirected and connected graph is a linear OSNI system. we show in the following that if an undirected and connected graph connects more than two identical linear OSNI systems, then the networked system also has an OSNI-like property. We construct a storage function for an $n$-node networked system based on the storage functions of all pairs of edge-linked nodes. Consider $n$ identical systems with transfer function $M(s)$ connected according to an undirected and connected graph $\mathcal{G}=(\mathcal{V},\mathcal{E})$. Let this networked system be defined by the transfer function matrix $\mathcal{M}_n(s)=\mathcal{L}_n\otimes M(s)$.
We define the storage function for the system which is a state-space realisation of $\mathcal{M}_n(s)$ as the sum of storage functions for all pairs of edge-linked nodes:
\begin{equation*}
    \hat V_2=\frac{1}{2}\sum_{(v_i,v_j)\in\mathcal{E}}V_2(\Delta x_{ij})=\frac{1}{2}\sum_{i,j\leq n}a_{ij}V_2(\Delta x_{ij}),
\end{equation*}
where $\mathcal{A}=[a_{ij}]\in \mathbb{R}^{n\times n}$ is the adjacency matrix of the graph $\mathcal{G}$.
The factor of $\frac{1}{2}$ is included because the summation counts the storage function of each pair of linked nodes $i$ and $j$ twice. According to (\ref{eq:dot V_2_{12}}), the time derivative of this storage function is
\begin{equation}\label{eq:OSNI-like property for networked OSNI system}
    \begin{aligned}
    \dot {\hat V}_2=&\ \frac{1}{2}\sum_{(v_i,v_j)\in\mathcal{E}}\dot V_2(\Delta x_{ij})\\
    \leq &\ \frac{1}{2}\sum_{(v_i,v_j)\in\mathcal{E}}\left(\Delta u_{ij}^T \Delta \dot y_{ij}-\delta |\Delta \dot y_{ij}|^2\right)\\
    =&\ U_2^T \dot Y_2-\frac{1}{2}\delta\sum_{(v_i,v_j)\in\mathcal{E}} |\Delta\dot y_{ij}|^2,
    \end{aligned}
\end{equation}
where $U_2$ and $Y_2$ are the input and output vectors of the system defined by $\mathcal{M}_n(s)$, respectively. Inequality (\ref{eq:OSNI-like property for networked OSNI system}) implies the system corresponding to $\mathcal{M}_n(s)$ satisfies the nonlinear NI definition. Moreover, there is an additional term $\sum_{(v_i,v_j)\in\mathcal{E}} |\Delta\dot y_{ij}|^2$ that describes the output strictness of the system. Comparing this term to $|\dot Y_2|^2$, which is expected to replace $\sum_{(v_i,v_j)\in\mathcal{E}} |\Delta\dot y_{ij}|^2$ in (\ref{eq:OSNI-like property for networked OSNI system}) if $\mathcal{M}_n(s)$ corresponds to a standard OSNI system, the term $\sum_{(v_i,v_j)\in\mathcal{E}} |\Delta\dot y_{ij}|^2$ gives a better measurement of the differences between the subsystem outputs in the networked systems. The term $\sum_{(v_i,v_j)\in\mathcal{E}} |\Delta\dot y_{ij}|^2$ only involves the output differences between pairs of nodes that are directly connected by an edge, while $|\dot Y_2|^2$ also involves the output differences of indirectly connected nodes, which is unnecessary. We say the system defined by $\mathcal{M}_n(s)$ is an OSNI-like system.
\subsection{Robust output feedback consensus of networked identical nonlinear NI systems}
\begin{figure}[h!]
\centering
\psfrag{H_0}{$\mathcal{H}_1$}
\psfrag{u_1}{$u_1$}
\psfrag{H_1}{$H_1$}
\psfrag{y_1}{$y_1$}
\psfrag{u_2}{$u_2$}
\psfrag{y_2}{$y_2$}
\psfrag{u_n}{$u_n$}
\psfrag{y_n}{$y_n$}
\psfrag{ddd}{\hspace{0.13cm}$\vdots$}
\psfrag{udd}{\hspace{0.1cm}$\vdots$}
\psfrag{odd}{\hspace{0.05cm}$\vdots$}
\includegraphics[width=8cm]{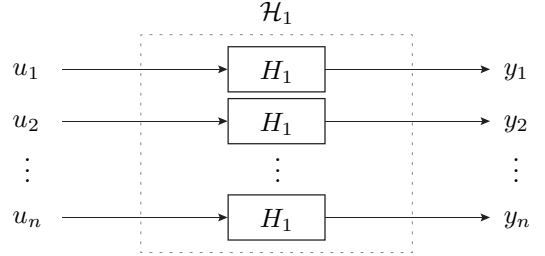}
\caption{System $\mathcal{H}_1$: a nonlinear system consisting of $n$ independent and identical nonlinear systems $H_1$, with independent inputs and outputs combined as the input and output of the networked system $\mathcal{H}_1$.}
\label{fig:H_1_multiple}
\end{figure}
Consider $n$ identical nonlinear systems $H_1$ described as in the state-space model (\ref{eq:state equation of nonlinear OSNI}), (\ref{eq:output equation of nonlinear OSNI}) which operate independently in parallel and each of them has its own input $u_i\in\mathbb{R}^m$ and output $y_i\in\mathbb{R}^m$, $(i=1, 2,...,n)$, as shown in Fig.~\ref{fig:H_1_multiple}. We combine their inputs and outputs respectively as the vectors
\begin{equation*}
    U_1=\left[\begin{matrix}u_1\\u_2\\\vdots\\u_n\end{matrix}\right]\in \mathbb R^{nm\times 1},\quad \textnormal{and} \quad Y_1=\left[\begin{matrix}y_1\\y_2\\\vdots\\y_n\end{matrix}\right]\in \mathbb R^{nm\times 1}.
\end{equation*}
We have the following lemma.
\begin{lemma}\label{lemma:NI property of networked NI}
If $H_1$ is a nonlinear NI system, then $\mathcal H_1$ is also a nonlinear NI system.
\end{lemma}
\begin{pf}
Since $H_1$ is a nonlinear NI system, there must exists a positive definite storage function $V_1(x)$ such that $\dot V_1(x)\leq u_i^T\dot y_i$. For the system $\mathcal H_1$, we define its storage function as $\hat V_1=\sum_{i=1}^n V_1(x_i)>0$. Then
\begin{equation}\label{eq:networked nonlinear NI property}
    \dot {\hat V}_1=\sum_{i=1}^n \dot V_1(x_i)\leq \sum_{i=1}^n u_i^T\dot y_i=U_1^T\dot Y_1,
\end{equation}
which implies the nonlinear NI inequality (\ref{eq:NI MIMO definition inequality}). Therefore, $\mathcal H_1$ is a nonlinear NI system.
\end{pf}
For a networked system $\mathcal{H}_1$ consisting of $n$ identical nonlinear NI subsystems $H_1$ with each subsystem described by the state-space model (\ref{eq:state equation of nonlinear OSNI}), (\ref{eq:output equation of nonlinear OSNI}), as shown in Fig.~\ref{fig:H_1_multiple}, output feedback consensus is defined as follows.
\begin{definition}
A distributed output feedback control law achieves output feedback consensus for a network of systems if $|y_i(t) - y_j(t)|\to 0$ as $t\to +\infty$, $\forall i,j\in\{1,2,\cdots,n\}$.
\end{definition}
\begin{figure}[h!]
\centering
\psfrag{H_0}{\hspace{-0.06cm}$\mathcal{H}_1$}
\psfrag{H_1}{\hspace{0.03cm}$H_1$}
\psfrag{U_1}{$U_1$}
\psfrag{Y_1}{\hspace{0.05cm}$Y_1$}
\psfrag{U_2}{$U_2$}
\psfrag{Y_2}{$Y_2$}
\psfrag{M}{$M(s)$}
\psfrag{M_s}{\hspace{-0.8cm}$\mathcal{M}_n(s)=\mathcal{L}_n\otimes M(s)$}
\psfrag{0}{$0$}
\psfrag{Lap}{\hspace{-0.2cm}$\mathcal{L}_n\otimes I_m$}
\psfrag{ddots}{\hspace{0.1cm}$\ddots$}
\psfrag{cdots}{\hspace{0.1cm}$\cdots$}
\psfrag{vdots}{\hspace{0.35cm}$\vdots$}
\includegraphics[width=8.5cm]{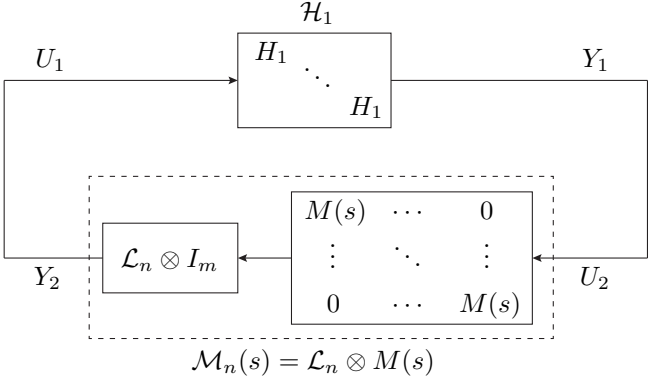}
\caption{Networked identical nonlinear NI systems with identical OSNI controllers connected according to a specified graph topology.}
\label{fig:closed_multiple}
\end{figure}

\begin{figure}[h!]
\centering
\psfrag{U_1}{$U_1$}
\psfrag{Y_1}{$Y_1$}
\psfrag{U_2}{$\ U_2$}
\psfrag{Y_2}{$Y_2$}
\psfrag{H_0}{$\mathcal{H}_1$}
\psfrag{M_n(s)}{\hspace{-0.3cm}$\mathcal{M}_n(s)$}
\includegraphics[width=8cm]{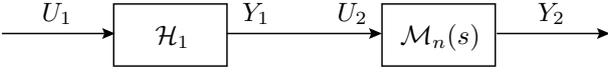}
\caption{Open-loop interconnection of the networked nonlinear NI system $\mathcal{H}_1$ and the networked linear OSNI system defined by $\mathcal{M}_n(s)$.}
\label{fig:open_multiple}
\end{figure}

Consider a network consisting of $n$ identical nonlinear NI systems $H_1$ with identical linear OSNI controllers defined by a transfer function matrix $M(s)$; e.g., see Fig.~\ref{fig:4_nodes} in the example. The entire system can be regarded as a closed-loop interconnection of two networked systems $\mathcal H_1$ and $\mathcal{M}_n(s)$, as shown in Fig. \ref{fig:closed_multiple}. According to Lemma \ref{lemma:NI property of networked NI}, the system $\mathcal H_1$ is a nonlinear NI system. Also, the system defined by $\mathcal M_n(s)$ has an OSNI-like property described in (\ref{eq:OSNI-like property for networked OSNI system}). The stability of the closed-loop interconnection shown in Fig.~\ref{fig:closed_multiple} is investigated in the sequel to prove the output feedback consensus of the networked nonlinear NI systems $H_1$ in $\mathcal H_1$.
First, we assume the following assumption is satisfied for the open-loop interconnection of the systems $\mathcal H_1$ and $\mathcal{M}_n(s)$ as shown in Fig.~\ref{fig:open_multiple}.
\\
Assumption IV: Given any constant input $U_1(t)\equiv\bar U_1$ for the system $\mathcal{H}_1$ and given its corresponding output $Y_1(t)$ (not necessarily constant) as input $U_2(t)$ to the system defined by $\mathcal{M}_n(s)$. If the corresponding output of the system defined by $\mathcal{M}_n(s)$ is constant; i.e., $Y_2(t)\equiv\bar Y_2$, then there exists a constant $\gamma\in(0,1)$ such that $\bar U_1$ and $\bar Y_2$ satisfy
\begin{equation}\label{eq:Assumption 4}
    \bar U_1^T \bar Y_2 \leq \gamma\left|\bar U_1\right|^2.
\end{equation}
Note that in most cases, a constant output $\bar Y_2$ can only be achieved when the system is in steady-state; i.e., when $U_2(t)\equiv\bar U_2$ and $Y_1(t)\equiv\bar Y_1$ are also constant. However, a special situation is also considered when a constant input $\bar U_1$ in the open-loop interconnection leads to a constant output $\bar Y_2$, while the system is not in steady-state, and $Y_1(t)$ and $U_2(t)$ still oscillate.
\begin{theorem}\label{theorem:consensus}
Given an undirected and connected graph $\mathcal{G}$ that models the communication links for networked identical nonlinear NI systems $H_1$, and given OSNI control law $M(s)$, robust output feedback consensus is achieved via the protocol
\begin{equation*}
    U_1=[\mathcal{L}_n\otimes M(s)]Y_1
\end{equation*}
as shown in Fig.~\ref{fig:closed_multiple}, if $H_1$ and $M(s)$ satisfy Assumptions I and II, $\mathcal{H}_1$ and $\mathcal{M}_n(s)$ satisfy Assumption IV, and the storage function defined as
\begin{equation*}
        \hat W:=\hat V_1+\hat V_2-Y_1^TY_2
\end{equation*}
is positive definite, where $\hat V_1$ and $\hat V_2$ are positive definite storage functions that satisfy the nonlinear NI property (\ref{eq:networked nonlinear NI property}) for system $\mathcal{H}_1$ and the OSNI-like property (\ref{eq:OSNI-like property for networked OSNI system}) for the system defined by $\mathcal{M}_n(s)$, respectively. Here $Y_1$ and $Y_2$ are outputs of the systems $\mathcal H_1$ and $\mathcal{M}_n(s)$, respectively.
\end{theorem}
\begin{pf}
We apply Lyapunov's direct method and take the time derivative of the storage function $\hat W$. According to (\ref{eq:networked nonlinear NI property}) and (\ref{eq:OSNI-like property for networked OSNI system}), we have
\begin{equation*}
\begin{aligned}
    \dot {\hat W}=&\ \dot {\hat V}_1+\dot {\hat V}_2-\dot Y_1^T Y_2-Y_1^T\dot Y_2\\
    =&\ \dot {\hat V}_1+\dot {\hat V}_2-U_1^T\dot Y_1-U_2^T\dot Y_2\\
    \leq& -\frac{1}{2}\delta\sum_{(v_i,v_j)\in\mathcal{E}} |\Delta\dot y_{ij}|^2\\
    \leq&\ 0.
\end{aligned}
\end{equation*}
This establishes the Lyapunov stability of this system. We now prove that output feedback consensus is achieved.

The Lyapunov derivative $\dot {\hat W}$ can only be zero when $\sum_{(v_i,v_j)\in\mathcal{E}} |\Delta\dot y_{ij}|^2=0$. This is equivalent to $\Delta \dot y_{ij}(t)=\dot y_i(t)-\dot y_j(t)=0$ for all $(v_i,v_j)\in\mathcal{E}$. In other words, $\dot W(X_1,X_2)$ cannot remain at zero unless $\dot y_i(t) \equiv \dot y_j(t)$ for all $(v_i,v_j)\in\mathcal{E}$. This means that there are always constant differences between the controller outputs $y_i(t)$ and $y_j(t)$ for all $(v_i,v_j)\in\mathcal{E}$; i.e., $\Delta y_{ij}(t)\equiv\overline {\Delta y_{ij}}$.

Recall the state-space model (\ref{eq:state-space difference state}), (\ref{eq:state-space difference output}) corresponds to a minimal realisation $(A,B,C)$ of $M(s)$. According to Assumptions I and II, $\Delta \dot y_{ij}(t)\equiv 0\implies \Delta \dot x_{ij}(t)\equiv 0\implies \Delta \dot u_{ij}(t)\equiv 0$. This implies constant differences between both the controllers' states $x_i(t), x_j(t)$ and the controllers' inputs $u_i(t), u_j(t)$, for all $(v_i,v_j)\in\mathcal{E}$; i.e., $\Delta x_{ij}(t)\equiv \overline{\Delta x_{ij}}$, $\Delta u_{ij}(t)\equiv \overline{\Delta u_{ij}}$. The state-space model (\ref{eq:state-space difference state}), (\ref{eq:state-space difference output}) can be modified to represent the constant differences between any pair of edge-linked controllers $i$ and $j$; i.e., $(v_i,v_j)\in\mathcal{E}$ implies
\begin{equation*}
\begin{aligned}
0=\Delta \dot x_{ij} = &\ A\overline{\Delta x_{ij}}+B\overline{\Delta u_{ij}},\\
\overline{\Delta y_{ij}} =&\ C\overline{\Delta x_{ij}}.\\
\end{aligned}
\end{equation*}
According to Definition~\ref{def:linear OSNI}, $A$ is Hurwitz and we can write
\begin{equation*}
    \overline{\Delta x_{ij}}= -A^{-1}B\overline{\Delta u_{ij}},
\end{equation*}
\begin{equation*}
  \overline{\Delta y_{ij}}=C\overline{\Delta x_{ij}}= -CA^{-1}B\overline{\Delta u_{ij}}=M(0)\overline{\Delta u_{ij}}.
\end{equation*}
The $i$-th $m\times 1$ vector in $Y_2$, which is also the distributed input to the $i$-th plant, can be expressed as
\begin{equation}\label{eq:Y_2_i}
\begin{aligned}
    &(Y_2)_i=\overline{(Y_2)_i}=\sum_{j=1}^na_{ij}\overline{\Delta y_{ij}}=\sum_{j=1}^na_{ij}M(0)\overline{\Delta u_{ij}}\\
    &=M(0)\sum_{j=1}^na_{ij}\overline{\Delta u_{ij}}=M(0)\overline{\left[(\mathcal{L}_n\otimes I_m) U_2\right]_i}.
\end{aligned}
\end{equation}
Therefore, the output $Y_2$ of the system defined by $\mathcal{M}_n(s)$ is
\begin{equation}\label{eq:Y_2,U_2 relation}
    Y_2=\overline{Y_2}=\left[I_n\otimes M(0)\right]\overline{(\mathcal{L}_n\otimes I_m)  U_2}=\left[\mathcal{L}_n\otimes M(0)\right] U_2.
\end{equation}
We now consider the closed-loop setting $\bar Y_2=\bar U_1$. Inequality (\ref{eq:Assumption 4}) implies
\begin{equation*}
    \bar U_1^T\bar Y_2=|\bar U_1|^2\leq\gamma|\bar U_1|^2,
\end{equation*}
which can only hold when $\bar Y_2=\bar U_1 = 0$. According to (\ref{eq:Y_2_i}), $\bar Y_2 \equiv 0\implies \overline{\Delta y_{ij}}\equiv 0\ \forall (v_i,v_j)\in\mathcal{E}$. This implies $\overline{\Delta x_{ij}}\equiv 0\implies \overline{\Delta u_{ij}}\equiv 0$ according to Assumptions I and II. $\overline{\Delta u_{ij}}\equiv 0$ $\forall (v_i,v_j)\in\mathcal{E}$ means the inputs of the controllers of any two edge-linked plants always have the same value, which means the outputs of the corresponding plants always have the same trajectory. Since the graph $\mathcal{G}$ is connected, this implies all plants have the same output trajectory. Hence output consensus is achieved. Otherwise, $\dot {\hat W}$ cannot remain at zero, and according to LaSalle's invariance principle, $\hat W$ will keep decreasing until either (i). output consensus is achieved; (ii). the states of all plants $H_1$ converge to zero, which also implies output consensus.
\end{pf}
\begin{remark}
The protocol in Theorem \ref{theorem:consensus} is robust against uncertainty in the system model for the subsystems connected in the network. For any network of identical nonlinear NI systems regardless of their model, we can apply this protocol to find a control law that enables the networked systems to achieve output feedback consensus.
\end{remark}
\section{Example}
In this section, we apply the output feedback consensus protocol developed in Section IV-C to a network of pendulum systems.

\begin{figure}[h!]
\centering
\psfrag{1}{\hspace{-0.02cm}\large$1$}
\psfrag{2}{\hspace{0.015cm}\large$2$}
\psfrag{3}{\hspace{-0.005cm}\large$3$}
\psfrag{4}{\hspace{0.01cm}\large$4$}
\includegraphics[width=4cm]{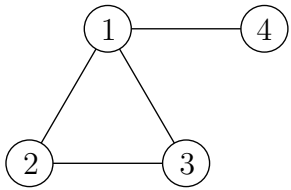}
\caption{An undirected and connected graph consisting of four nodes.}
\label{fig:4_nodes}
\end{figure}

Consider a simple networked system consisting of four identical pendulum systems connected by a graph $\mathcal{G}$ as shown in Fig.~\ref{fig:4_nodes}. The Laplacian matrix of the graph $\mathcal G$ is $$\mathcal{L}_4=\left[\begin{matrix}3&-1&-1&-1\\-1&2&-1&0\\-1&-1&2&0\\-1&0&0&1\\ \end{matrix}\right].$$

\begin{figure}[h!]
\centering
\psfrag{Output theta}{Output}
\psfrag{time (s)}{Time (s)}
\psfrag{Output Feedback consensus of pendulums}{\hspace{-0.5cm}Output Feedback Consensus of Pendulums}
\psfrag{Pendulum sys 1}{\small Pendulum 1}
\psfrag{Pendulum sys 2}{\small Pendulum 2}
\psfrag{Pendulum 3}{\small Pendulum 3}
\psfrag{Pendulum 4}{\small Pendulum 4}
\psfrag{-3}{-3}
\psfrag{-2}{-2}
\psfrag{-1}{-1}
\psfrag{0}{0}
\psfrag{1}{1}
\psfrag{2}{2}
\psfrag{3}{3}
\psfrag{0}{0}
\psfrag{5}{5}
\psfrag{15}{15}
\psfrag{10}{10}
\psfrag{20}{20}
\includegraphics[width=9cm]{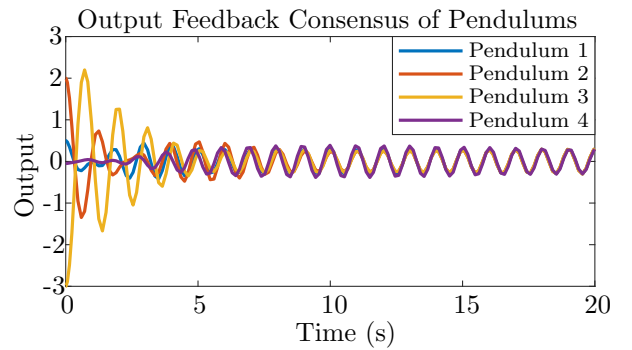}
\caption{Robust output feedback consensus for networked pendulum systems.}
\label{fig:sim1}
\end{figure}

The pendulum has a state-space model
\begin{equation*}
\begin{aligned}
    \left[\begin{matrix}\dot x_1\\\dot x_2\end{matrix}\right]=&\left[\begin{matrix} x_2\\ \frac{1}{ml^2}\left(-\kappa x_1-mgl\sin {x_1}+u_1\right)\end{matrix}\right],\\
    y_1=&\ x_1,
\end{aligned}
\end{equation*}
where $m=1kg$ is the mass of the bob, $l=0.5m$ is the length of the rod, $\kappa=5N\cdot m/rad$ is the spring constant of a torsional spring installed in the pivot and $g\approx 9.8m/s^2$ is the gravitational acceleration. The system states are $x_1$ the counterclockwise angular displacement from the vertically downward position and $x_2$ is the system angular velocity. The system's input $u$ is an external torsional force to the counterclockwise direction, and $y$ is the system's output. The pendulum plant is not passive systems hence the existing passivity-based consensus methods is not applicable. In contrast, the consensus result proposed in this paper is effective because the pendulum plant is a nonlinear NI system with the storage function $V_1(x_1,x_2)=\frac{1}{2}\kappa x_1^2+\frac{1}{2}ml^2x_2^2+mgl(1-\cos{x_1})$. We apply a networked OSNI controller defined by $M(s)=\frac{a}{s+b}$ according to Fig.~\ref{fig:closed_multiple}. A minimal realisation of $M(s)$ is
\begin{equation*}
\begin{aligned}
    \dot x_3=&-bx_3+au_2,\\
    y_2=&\ x_3.
\end{aligned}
\end{equation*}
The controller is an OSNI system with the storage function $V_2(x_3)=\frac{b}{2a}x_3^2$. According to Theorem \ref{theorem:consensus}, we use the protocol $U_1=(\mathcal{L}_4\otimes M(s))Y_1$ to control the pendulum plants. We choose $a=10$ and $b=10$ as the controller parameters, which will guarantee the positive definiteness of the storage function of the entire networked system. Assumptions I, II and IV are also verified. All of the pendulums in this graph tend to converge to the same trajectory, which is shown in Fig.~\ref{fig:sim1}.

It is demonstrated in this example that the protocol presented in this paper is an alternative approach to achieve output feedback consensus when passivity-based consensus approaches are not applicable.
\section{Conclusion}
The robust output feedback consensus problem is investigated in this paper for networked identical nonlinear NI systems. To obtain more generality, the definition of nonlinear NI systems is extended to MIMO systems and the definition of OSNI systems is extended to include nonlinear systems. The closed-loop interconnection of a nonlinear NI and a nonlinear OSNI system is proved to be asymptotically stable. The nonlinear NI property and an OSNI-like property is proved for networked identical nonlinear NI systems and networked identical linear OSNI systems, respectively. These properties are then applied to analyse the stability of the closed-loop interconnection of networked nonlinear NI systems and networked linear OSNI systems, which proves robust output feedback consensus for networked identical nonlinear NI plants.

\section{Succeeding Results}\label{sec:succeeding}
Due to the delay of publication of this paper caused by the Covid19 pandemic, several succeeding results have already been published before the publication of this paper. Output feedback consensus problem is addressed for heterogeneous nonlinear NI systems in \cite{shi2020robustb}, which covers the problem investigated in this paper. The definition of nonlinear NI systems is extended in \cite{shi2020robustc} to allow for systems with free body motion. Nonlinear OSNI systems are redefined as well. A stability result is established for the interconnection of a single nonlinear NI system and a single nonlinear OSNI system. Then a control framework is proposed to achieve output feedback consensus for networked heterogeneous nonlinear NI systems with free body motion.

\begin{ack}
The author K. Shi would like to thank Dr. Ahmed G. Ghallab for helpful discussions in the context of \cite{ghallab2018extending}.
\end{ack}


\begin{thebibliography}{18}
\providecommand{\natexlab}[1]{#1}
\providecommand{\url}[1]{\texttt{#1}}
\providecommand{\urlprefix}{URL }
\expandafter\ifx\csname urlstyle\endcsname\relax
  \providecommand{\doi}[1]{doi:\discretionary{}{}{}#1}\else
  \providecommand{\doi}{doi:\discretionary{}{}{}\begingroup
  \urlstyle{rm}\Url}\fi

\bibitem[{Bhikkaji et~al.(2011)Bhikkaji, Moheimani, and
  Petersen}]{bhikkaji2011negative}
Bhikkaji, B., Moheimani, S.R., and Petersen, I.R. (2011).
\newblock A negative imaginary approach to modeling and control of a collocated
  structure.
\newblock \emph{IEEE/ASME Transactions on Mechatronics}, 17(4), 717--727.

\bibitem[{Bhowmick and Lanzon(2019)}]{bhowmick2019output}
Bhowmick, P. and Lanzon, A. (2019).
\newblock Output strictly negative imaginary systems and its connections to
  dissipativity theory.
\newblock In \emph{2019 IEEE 58th Conference on Decision and Control (CDC)},
  6754--6759. IEEE.

\bibitem[{Bhowmick and Patra(2017)}]{bhowmick2017lti}
Bhowmick, P. and Patra, S. (2017).
\newblock On {LTI} output strictly negative-imaginary systems.
\newblock \emph{Systems $\&$ Control Letters}, 100, 32--42.

\bibitem[{Cai and Hagen(2010)}]{cai2010stability}
Cai, C. and Hagen, G. (2010).
\newblock Stability analysis for a string of coupled stable subsystems with
  negative imaginary frequency response.
\newblock \emph{IEEE Transactions on Automatic Control}, 55(8), 1958--1963.

\bibitem[{Das et~al.(2014{\natexlab{a}})Das, Pota, and Petersen}]{das2014mimo}
Das, S.K., Pota, H.R., and Petersen, I.R. (2014{\natexlab{a}}).
\newblock A {MIMO} double resonant controller design for nanopositioners.
\newblock \emph{IEEE Transactions on Nanotechnology}, 14(2), 224--237.

\bibitem[{Das et~al.(2014{\natexlab{b}})Das, Pota, and
  Petersen}]{das2014resonant}
Das, S.K., Pota, H.R., and Petersen, I.R. (2014{\natexlab{b}}).
\newblock Resonant controller design for a piezoelectric tube scanner: A mixed
  negative-imaginary and small-gain approach.
\newblock \emph{IEEE Transactions on Control Systems Technology}, 22(5),
  1899--1906.

\bibitem[{Das et~al.(2015)Das, Pota, and Petersen}]{das2015multivariable}
Das, S.K., Pota, H.R., and Petersen, I.R. (2015).
\newblock Multivariable negative-imaginary controller design for damping and
  cross coupling reduction of nanopositioners: a reference model matching
  approach.
\newblock \emph{IEEE/ASME Transactions on Mechatronics}, 20(6), 3123--3134.

\bibitem[{Ghallab et~al.(2018)Ghallab, Mabrok, and
  Petersen}]{ghallab2018extending}
Ghallab, A.G., Mabrok, M.A., and Petersen, I.R. (2018).
\newblock Extending negative imaginary systems theory to nonlinear systems.
\newblock In \emph{2018 IEEE Conference on Decision and Control (CDC)},
  2348--2353. IEEE.

\bibitem[{Lanzon and Petersen(2008)}]{lanzon2008stability}
Lanzon, A. and Petersen, I.R. (2008).
\newblock Stability robustness of a feedback interconnection of systems with
  negative imaginary frequency response.
\newblock \emph{IEEE Transactions on Automatic Control}, 53(4), 1042--1046.

\bibitem[{Mabrok et~al.(2014)Mabrok, Kallapur, Petersen, and
  Lanzon}]{mabrok2014generalizing}
Mabrok, M.A., Kallapur, A.G., Petersen, I.R., and Lanzon, A. (2014).
\newblock Generalizing negative imaginary systems theory to include free body
  dynamics: Control of highly resonant structures with free body motion.
\newblock \emph{IEEE Transactions on Automatic Control}, 59(10), 2692--2707.

\bibitem[{Patra and Lanzon(2011)}]{patra2011stability}
Patra, S. and Lanzon, A. (2011).
\newblock Stability analysis of interconnected systems with “mixed”
  negative-imaginary and small-gain properties.
\newblock \emph{IEEE Transactions on Automatic Control}, 56(6), 1395--1400.

\bibitem[{Petersen and Lanzon(2010)}]{petersen2010feedback}
Petersen, I.R. and Lanzon, A. (2010).
\newblock Feedback control of negative-imaginary systems.
\newblock \emph{IEEE Control Systems Magazine}, 30(5), 54--72.

\bibitem[{{Shi} et~al.(2020){Shi}, {Vladimirov}, and
  {Petersen}}]{shi2020robustb}
{Shi}, K., {Vladimirov}, I.G., and {Petersen}, I.R. (2020).
\newblock Robust output feedback consensus for networked heterogeneous
  nonlinear negative-imaginary systems.
\newblock In \emph{2020 Australian and New Zealand Control Conference (ANZCC)},
  214--219.
\newblock \doi{10.1109/ANZCC50923.2020.9318395}.

\bibitem[{Shi et~al.(2021{\natexlab{a}})Shi, Petersen, and
  Vladimirov}]{shi2021negative}
Shi, K., Petersen, I.R., and Vladimirov, I.G. (2021{\natexlab{a}}).
\newblock Negative imaginary state feedback equivalence for systems of relative
  degree one and relative degree two.
\newblock \emph{Submitted to 2021 IEEE 60th Conference on Decision and Control
  (CDC), arXiv preprint arXiv:2103.05249}.

\bibitem[{Shi et~al.(2021{\natexlab{b}})Shi, Petersen, and
  Vladimirov}]{shi2020robustc}
Shi, K., Petersen, I.R., and Vladimirov, I.G. (2021{\natexlab{b}}).
\newblock Output feedback consensus for networked heterogeneous nonlinear
  negative-imaginary systems with free body motion.
\newblock \emph{Submitted to IEEE Transactions on Automatic Control, arXiv
  preprint arXiv:2011.14610}.

\bibitem[{Song et~al.(2012)Song, Lanzon, Patra, and
  Petersen}]{song2012negative}
Song, Z., Lanzon, A., Patra, S., and Petersen, I.R. (2012).
\newblock A negative-imaginary lemma without minimality assumptions and robust
  state-feedback synthesis for uncertain negative-imaginary systems.
\newblock \emph{Systems \& Control Letters}, 61(12), 1269--1276.

\bibitem[{Wang et~al.(2015)Wang, Lanzon, and Petersen}]{wang2015robust}
Wang, J., Lanzon, A., and Petersen, I.R. (2015).
\newblock Robust output feedback consensus for networked negative-imaginary
  systems.
\newblock \emph{IEEE Transactions on Automatic Control}, 60(9), 2547--2552.

\bibitem[{Xiong et~al.(2010)Xiong, Petersen, and Lanzon}]{xiong2010negative}
Xiong, J., Petersen, I.R., and Lanzon, A. (2010).
\newblock A negative imaginary lemma and the stability of interconnections of
  linear negative imaginary systems.
\newblock \emph{IEEE Transactions on Automatic Control}, 55(10), 2342--2347.

\end{thebibliography}
\end{document}